\documentclass[preprint]{aastex}

\newcommand{\be}{\begin{equation}}
\newcommand{\ee}{\end{equation}}
\newcommand{\ebar}{{\langle e \rangle}} 
 
\newcommand{\albar}{{\bar \alpha}} 
\newcommand{\eigenv}{{ \Lambda }} 
\newcommand{\disturb}{{ {\cal V} }} 
\newcommand{\taugr}{{ \tau_{\rm gr} }} 
\newcommand\sgn{ {\rm sign}} 
\def\lta{\,\raise 0.3 ex\hbox{$ < $}\kern -0.75 em
 \lower 0.7 ex\hbox{$\sim$}\,}
\def\gta{\,\raise 0.3 ex\hbox{$ > $}\kern -0.75 em
 \lower 0.7 ex\hbox{$\sim$}\,} 
\newcommand\pp{\parshape 2 0.0truecm 15.5truecm 1.25truecm 14.25truecm} 

\begin{document}
\baselineskip=20pt 
 
\title{RELATIVISTIC EFFECTS IN EXTRASOLAR PLANETARY SYSTEMS} 

\author{\bf Fred C. Adams$^{1,2}$ and Gregory Laughlin$^3$ } 

\affil{$^1$Michigan Center for Theoretical Physics \\
Physics Department, University of Michigan, Ann Arbor, MI 48109} 

\affil{$^2$Astronomy Department, University of Michigan, Ann Arbor, MI 48109}

\affil{$^3$Lick Observatory, University of California, Santa Cruz, CA 95064} 

\email{fca@umich.edu}

\begin{abstract} 
\baselineskip=20pt 

This paper considers general relativistic (GR) effects in currently
observed extrasolar planetary systems. Although GR corrections are
small, they can compete with secular interactions in these systems and
thereby play an important role. Specifically, some of the observed
multiple planet systems are close to secular resonance, where the
dynamics is extremely sensitive to GR corrections, and these systems
can be used as laboratories to test general relativity. For the
three-planet solar system Upsilon Andromedae, secular interaction
theory implies an 80\% probability of finding the system with its
observed orbital elements if GR is correct, compared with only a 2\%
probability in the absence of GR. In the future, tighter constraints
can be obtained with increased temporal coverage.

\end{abstract}

\bigskip 
$\,$ 
\bigskip 

{This essay received ``Honorable Mention'' in the 2006 Essay
Competition of the Gravity Research Foundation }

\newpage 

\noindent 
{\bf 1. Introduction} 

\noindent 
The perihelion advance of the planet Mercury is one of the classic
tests of general relativity [1].  The recent discoveries of extrasolar
planets [2] provide us with a new ensemble of solar systems to study
periastron advance and other relativistic effects. The current sample
of extrasolar planets [3] includes many orbits that are surprisingly
close to the stars, with semimajor axes $a \approx 0.05$ AU and hence
periods $P \approx 4$ days. Such small semimajor axes imply that these
planets experience larger GR effects than the planets in our solar
system. One way to quantify the efficacy of GR is through the
dimensionless parameter $\mu \equiv GM_\ast/(c^2 a)$. A planet in a 4
day orbit has a $\mu$ parameter nearly 10 times larger than that of
Mercury, but direct measurement of periastron advance remains
difficult because of the large number of observations required and
because close planets tend to have nearly circular
orbits. Fortunately, however, planetary systems with multiple planets
and particular architectures allow for GR to exhibit much more
pronounced effects. In these systems, secular interactions between the
planets enforce time variations in the orbital elements (e.g.,
eccentricity $e$). These interactions are sensitive to the exact
tuning of the system into secular resonance, and such tuning is
affected by the relativistic corrections to the classical theory. As a
result, these systems provide a new test of general relativity. In
other systems, we find that this sensitivity to GR allows us to place
new constraints on the system parameters.

\noindent 
{\bf 2. Theory of Secular Interactions}  

\noindent
We first outline the basic theory of secular interactions in
multi-planet solar systems. In the absence of relativistic
corrections, this topic has been widely discussed previously [4]. The
basic result of these interactions is to force the orbits of
participating planets to exchange angular momentum and thereby display
time varying eccentricities (and other orbital elements). The
amplitudes and time scales of these variations depend on the solar
system configuration.  Here we briefly outline the formalism and add
the leading order relativistic correction. To the second order in
eccentricity and inclination angle, the equations of motion for
eccentricity $e_j$ and argument of periastron $\varpi_j$ decouple from
those of inclination angle and the ascending node. Following standard
convention [4], we work in terms of the variables defined by
\be
h_j \equiv e_j \sin \varpi_j \,  \qquad {\rm and} \qquad 
k_j \equiv e_j \cos \varpi_j \, , 
\ee 
where the subscript refers to the $j$th planet in a solar system 
with $N$ planets. The equations of motion take the form 
\be 
{d h_j \over dt} = {1 \over n_j a_j^2} 
{\partial \disturb_j \over \partial k_j} 
\qquad {\rm and} \qquad 
{d k_j \over dt} = - {1 \over n_j a_j^2} 
{\partial \disturb_j \over \partial h_j} \, , 
\ee 
where $\disturb_j$ the secular part of the disturbing function, $n_j$
is the mean motion, and $a_j$ is the semi-major axis (for the $j$th
planet).  To consistent order in this approximation, the disturbing
function can be written
\be 
\disturb_j = n_j a_j^2 \Bigl[ {1 \over 2} A_{jj} e_j^2 + 
\sum_{k \ne j} A_{jk} e_j e_k \cos (\varpi_j - \varpi_k) 
\Bigr] \, . 
\ee 
The physics of secular interactions is thus encapsulated in the 
$N \times N$ matrix $A_{ij}$, where the matrix elements take the form 
\be 
A_{jj} = n_j \Bigl[ {1 \over 4} \sum_{k \ne j} {m_k \over M_\ast + m_j} 
\alpha_{jk} \albar_{jk} b^{(1)}_{3/2}(\alpha_{jk}) \, + 
3 {G M_\ast \over c^2 a_j}  \Bigr] \, , 
\label{eq:diag} 
\ee 
\be
A_{jk} = - n_j {1 \over 4} {m_k \over M_\ast + m_j} 
\alpha_{jk} \albar_{jk} b^{(2)}_{3/2}(\alpha_{jk}) \, .
\label{eq:offdiag} 
\ee
The quantities $\alpha_{jk}$ are defined such that $\alpha_{jk} =
a_j/a_k$ ($a_k/a_j$) if $a_j < a_k$ ($a_k < a_j$). The complementary
quantities $\albar_{jk}$ are defined so that $\albar_{jk} = a_j/a_k =
\alpha_{jk}$ if $a_j < a_k$, but $\albar_{jk}$ = 1 for $a_k < a_j$.
Finally, $b^{(1)}_{3/2}(\alpha_{jk})$ and $b^{(2)}_{3/2}(\alpha_{jk})$
are the Laplace coefficients [4]. The diagonal matrix elements
include the leading order correction for general relativity [5]. These
corrections are small in an absolute sense, with $\mu \sim 4 \times
10^{-6} (a_j/0.05 {\rm AU})^{-1}$, but they compete with the other 
terms and affect the eigenfrequencies when the system is near
resonance.  With the above definitions, the time variations of the
eccentricity and argument of periastron are given by
\be 
h_j (t) = \sum_i \eigenv_{ji} \sin (\lambda_i t + \beta_i)\, , \qquad 
k_j (t) = \sum_i \eigenv_{ji} \cos (\lambda_i t + \beta_i)\, , 
\ee
where the $\lambda_i$ are eigenvalues of the matrix $A_{ij}$ 
and the $\eigenv_{ji}$ are the corresponding eigenvectors. The phases 
$\beta_i$ and the normalization of the eigenvectors are determined 
by the initial conditions, i.e., the values of eccentricity $e_j$ 
and argument of periastron $\varpi_j$ for each planet at $t=0$
(taken to be the time when the orbital elements of the extrasolar 
planets are measured).  

\noindent
{\bf 3. General Relativity in Observed Systems}  

\noindent
To illustrate the action of GR in extrasolar planetary systems, we
have calculated the secular interactions for two observed systems [3]
with and without including the general relativistic terms. We use the
systems Upsilon Andromedae and HD160691 because they contain inner
planets with $a \sim 0.05$ AU and outer planets with $a \sim 1$ AU. As
shown below, this type of solar system architecture allows for GR to
compete with secular interactions. Figure 1 shows the mean
eccentricity $\ebar$ of the innermost planet, as driven by secular
interactions and averaged over many cycles, plotted as a function of
$\sin i$ for the two systems.

For Upsilon Andromedae (top panel), general relativity acts to damp
the excitation of eccentricity by secular interactions. Since the
viewing angle is not measured, the resulting mean eccentricity $\ebar$
is shown as a function of $\sin i$. For large $\sin i$ values, the
inner planet would be driven to mean eccentricity values $\ebar
\approx 0.4$ without GR, but only $\ebar \approx 0.016$ when
relativistic corrections are included. The observed eccentricity for
the inner planet $e_{obs} \approx 0.011$ is much closer to the
relativistic mean value. Since secular theory uses the observed
orbital elements in the boundary conditions, the system always has
some chance of displaying the observed (low) eccentricity of the inner
planet (even if $\ebar \sim 0.4$), so the implications for GR must be
stated in terms of probabilities: If the observed eccentricity of the
inner planet has a measurement error that is gaussian distributed with
width $\sigma_{obs} = 0.015$, then the probability of observing the
system in its current state would be only 0.023 in the absence of GR.
The probability of finding the inner planet with its observed
eccentricity is 0.78 if GR is included. The validity of general
relativity is thus strongly favored.

For the HD160691 system (bottom panel), the inclusion of relativistic
terms acts in the opposite direction, i.e., it leads to greater
predicted values of $\ebar$. For values of $\sin i \sim 1$, the
predicted mean driven eccentricities are small enough to be consistent
with the observed low value $e_{obs} \sim 0$. As $\sin i$ decreases,
however, the level of eccentricity forcing increases and the system
reaches a resonance near $\sin i \approx 0.5$. The observed low value
of eccentricity, in conjunction with these considerations, thus
constrain the viewing angle of the HD160691 system to be nearly
edge-on; if we require $\ebar \lta 0.05$, consistent with
observational uncertainties, the viewing angle is confined to the
range $\sin i \gta 0.93$ ($i \gta 70^\circ$). Since inclination
angles are notoriously difficult to measure in these systems, this
constraint on $i$ is quite valuable. For example, this limit on the
viewing angle has important implications for the possibility of 
observing the inner planet in transit [6]. 

\noindent
{\bf 4. The Magnitude of Relativistic Effects}  

\noindent
Next we want to find an analytic criterion that characterizes the size
of relativistic effects. In these systems, general relativity is
significant when the final term in equation (\ref{eq:diag}) competes
with the others.  In practice, only the inner planet has significant
relativistic corrections. We can also assume that the system is
sufficiently hierarchical so that $b^{(1)}_{3/2}(\alpha_{j1}) = 3
\alpha_{j1} + {\cal O} (\alpha_{j1}^2)$, and that only one outer
planet competes with the relativistic correction. In this limit, we
obtain the requirement
\be 
{m_j \over M_\ast} \alpha_{j1}^3 \approx {4 G M_\ast \over c^2 a_1} 
\, , 
\ee 
where the subscript `1' denotes the inner planet and `$j$' the outer
planet. Both sides of this equation represent small dimensionless 
quantities. When their ratio is of order unity, however, relativistic
effects compete with secular interactions, i.e., this constraint is
equivalent to the requirement that one of the dimensionless fields
$\Pi \sim 1$, where 
\be 
\Pi \equiv {4 G M_\ast^2 a_j^3 \over c^2 m_j a_1^4 } \, \approx 
6.3 \Bigl( {m_j \over m_J} \Bigr)^{-1} 
\Bigl( {M_\ast \over M_\odot} \Bigr)^{2} 
\Bigl( {a_j \over 1 \, {\rm AU} } \Bigr)^{3} 
\Bigl( {a_1 \over 0.05 \, {\rm AU}} \Bigr)^{-4} \, . 
\label{eq:pidef} 
\ee  
The second equality indicates that relativistic effects compete with
secular interactions for a Jovian planet in a $\sim$1 AU orbit
perturbing an inner planet in a $\sim 0.05$ AU orbit. The relative
size of the relativistic effect grows with increasing semi-major axis
$a_j$ of the perturber, but the absolute size of the secular
effect decreases. For closer perturbing planets ($a_j \ll 1$ AU),
interactions are stronger but relativity plays little role; for more
distant planets ($a_j \gg 1$ AU), the interactions are weak but are
dominated by relativity, which only makes the inner planet precess
forward in its orbit. The condition for GR to compete with secular 
interactions can also be written in terms of time scales. GR itself 
defines a characteristic time scale $\taugr$, 
\be
\taugr \equiv {c^2 a_1^{3/2} \over 3 (G M_\ast)^{3/2}} \approx 3011 \, 
{\rm yr} \, \Bigl( {a_1 \over 0.05 \, {\rm AU} } \Bigr)^{5/2} \, 
\Bigl( {M_\ast \over 1.0 M_\odot } \Bigr)^{-3/2} \, , 
\ee 
the time required for the periastron to precess one radian forward in
its orbit. When this time scale is comparable to the secular
interaction time scale, $\taugr \sim \tau_{\rm sec}$, then GR plays a
significant role in the dynamics. For planets with $\sim4$ day orbits,
$\taugr$ is comparable to the secular interaction time scales for many
of the observed extrasolar planetary systems, where $\tau_{\rm sec}$ =
$10^2 - 10^5$ yr [6]. 

\noindent
{\bf 5. The Sign of Relativistic Effects} 

\noindent
Figure 1 shows that GR can lead to either larger (HD160691) or smaller
(Ups And) eccentricity forcing, compared with classical
theory. However, most previous discussions of relativistic effects in
planetary systems emphasize its stabilizing influence -- the tendency
for GR to reduce the amplitude of eccentricity forcing. It is thus
useful to explore when the two types of behavior occur, and what
system parameters are required.

For a two planet system, we consider an idealized case in which the
inner planet has a nearly circular orbit, or at least cycles through
the $e$ = 0 state. This condition is common in that close planets in
observed systems tend to have nearly circular orbits [3]. In this
limit, the eccentricity amplitude $\eta$ of the inner planet (forced
by the outer planet) can be written in the form 
\be
\eta^2 = {A_{12}^2 e_{2(0)}^2 \over (\lambda_1 - \lambda_2)^2 } \, ,   
\ee
where $e_{2(0)}$ is the initial eccentricity of the outer planet. For
a two planet system, the difference in eigenvalues is positive
definite and is given by the expression
\be 
\Delta \lambda = \lambda_1 - \lambda_2 =  
\bigl[ (A_{11} - A_{22})^2 + 4 A_{12} A_{21} \bigr]^{1/2} \, . 
\ee 
Since the eigenvalues cannot be degenerate for a two planet system,
secular interactions do not lead to resonance.  Notice that only the
diagonal matrix elements contain GR corrections and only the inner
planet is close enough to the star for such corrections to matter.
The first matrix element can thus be written $A_{11} = A_0 + 3 n_1
\mu$, where $A_0$ is the matrix element in the absence of GR and 
$\mu = G M_\ast/(c^2 a_1)$ is the relativistic correction factor.
With this construction, the question of whether GR acts to increase 
or decrease the amplitude of eccentricity oscillation depends on
the sign of the derivative $d \eta / d\mu$, i.e.,
\be
\sgn \Big( {d \eta \over d \mu} \Bigr) = - 
\sgn (A_0 + 3 n_1 \mu - A_{22}) = - 
\sgn \Big[ 1 + \Pi - {m_1 \over m_2} 
\big( {a_1 \over a_2} \bigr)^{1/2} \Big] \, . 
\ee 
The second equality uses the approximations $m_1, m_2 \ll M_\ast$,
$a_1 \ll a_2$, and $b_{3/2}^{(1)} \approx 3 a_1/a_2$; the
dimensionless field $\Pi$ is defined in equation (\ref{eq:pidef}). The
third term must dominate in order for the sign to be positive, and
hence for relativity to lead to greater eccentricity amplitudes.
Since $a_1 < a_2$ by definition, relativity will lead to greater
stability (smaller amplitudes) unless $m_1 > m_2$. As a result, 
relativity can amplify eccentricity forcing only for sufficiently 
massive inner planets, where the requirement for amplification 
can be written in the form
\be 
m_1 \, > \, m_2 \, \big( {a_2 \over a_1} \bigr)^{1/2} 
+ \, 4 \mu M_\ast \big( {a_2 \over a_1} \bigr)^{7/2} \, . 
\label{eq:mrellimit} 
\ee 
This constraint thus implies that the inner planet mass $m_1$ must be
large compared to the outer planet mass $m_2$. The two planet systems
observed to date [3] show an interesting trend: None of these solar
systems have planetary masses that satisfy this inequality, and hence
GR leads to greater stability in all of these (two planet) cases.

Now we consider a three planet system in which the inner planet has
little effect on the outer two planets due to its smaller mass and/or
inner position. These characteristics apply to both HD160691 and
UpsAnd. In this hierarchical limit, the $3 \times 3$ matrix $A_{ij}$
can be approximated by taking $A_{12} = A_{13} = A_{21} = A_{31} = 0$. 
The eigenvalues of this reduced matrix then take the form
\be
\lambda_1 = A_{11} \, , 
\lambda_{2,3} = {1 \over 2} \bigl\{ (A_{22} + A_{33} \pm 
\bigl[ (A_{33} - A_{22})^2 + 4 A_{23} A_{32} \bigr]^{1/2} 
\bigr\} \, . 
\ee 
With this level of complexity, the system can have degenerate
eigenvalues, e.g., when $\lambda_1 = A_{11}$ is equal to either
$\lambda_2$ or $\lambda_3$.  Without loss of generality, suppose that
$\lambda_1$ and $\lambda_2$ are nearly degenerate. The effect of GR is
to add a small positive contribution to $A_{11}$ and hence $\lambda_1$
(we again take $A_{11} = A_0 + 3 n_1 \mu$), where the added term
describes the relativistic precession of the inner planet's orbit.  
If $\lambda_1 \lta \lambda_2$, then GR brings the eigenvalues closer 
together and thus acts to increase eccentricity forcing. If $\lambda_1
\gta \lambda_2$, then GR makes the eigenvalues more unequal and acts
to decrease eccentricity forcing.  The full cubic equation for $\det
[A - \lambda I]$ is more complicated, but allows for similar behavior.

\noindent
{\bf 6. Conclusion} 

\noindent
For multiple planet systems with favorable architectures, the effects
of general relativity can be significant (Fig. 1). When solar systems
have two relatively massive outer planets and a third inner planet of
smaller mass near secular resonance, GR effects are large enough to
move the system in or out of a resonant condition.  Since the
resonance condition depends on planetary masses, which in turn depend
on $\sin i$, and since small planets cannot survive in exact
resonance, this effect can be used to constrain the allowed range of
$\sin i$ in observed extrasolar planetary systems.  For the HD160691
system, this constraint implies that $\sin i \gta 0.93$.  This GR
effect thus provides another means of probing the properties of these
solar systems. For other systems, general relativity causes the inner
planet to precess forward in its orbit fast enough to compromise
eccentricity excitation from a second planet, i.e., the mean
eccentricity values are much smaller than they would be in flat
space. This latter effect can be used as a new test of general
relativity. For example, the Upsilon Andromedae system has an
$\sim80\%$ chance of being observed with its measured parameters if GR
is correct, but only a $\sim2\%$ chance if the GR corrections were
absent.  Although the weak-field limit of GR (including the perihelion
advance of Mercury) is well established in our own solar system [1],
the results of this paper add to our understanding by: [A] Developing
a new relativistic effect -- with amplified sensitivity -- due to GR
competing with secular interactions, and [B] Providing an independent
confirmation of GR in other solar systems.  In the future, additional
extrasolar planetary systems and/or higher precision observational
data can provide more stringent tests of general relativity.

$\,$  
\bigskip 

{\bf References} 

\par\pp 1. 
C. M. Will, {\sl Theory and Experiment in Gravitational Physics} 
(Cambridge Univ. Press, Cambridge, 1993). 

\par\pp 2. 
G. Marcy and P. R. Butler, P. R., Astrophys. J. {\bf 464}, L147 (1996); 
M. Mayor and D. Queloz, Nature, {\bf 378}, 355 (1995). 

\par\pp 3. 
California and Carnegie Planet Search, http://exoplanets.org/

\par\pp 4. 
C. D. Murray and S. F. Dermott, {\sl Solar System Dynamics} 
(Cambridge Univ. Press, Cambridge, 1999). 

\par\pp 5. 
S. Weinberg, {\sl Gravitation and Cosmology} (Wiley, New York, 1972).  

\par\pp 6. 
F. C. Adams and G. Laughlin, Astrophys. J., {\bf 649}, 992 (2006). 

\newpage 
\begin{figure}
\figurenum{1}  
\plotone{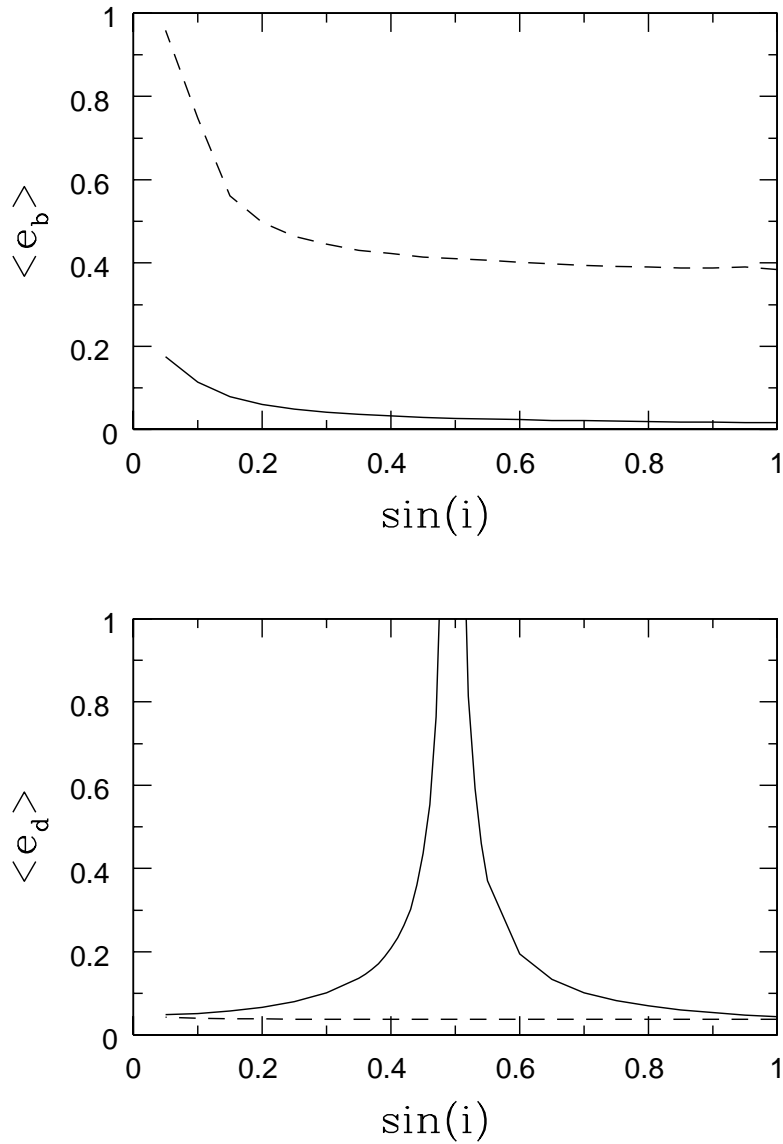} 
\caption{\small The effects of general relativistic corrections on two
extrasolar planetary systems. The mean eccentricity $\ebar$ of the
innermost planet, as driven by secular interactions and averaged over
many cycles, is plotted as a function of $\sin i$ for the Upsilon
Andromedae system (top panel) and the HD160691 system (bottom
panel). Both systems have three planets detected to date. The
predictions of secular theory are shown with relativistic corrections
as the solid curves, and without relativistic corrections as the
dashed curves. Notice that the relativistic terms act in opposite ways
in the two systems: Inclusion of relativity acts to damp eccentricity
excitation by the secular interactions in the Upsilon Andromedae system. 
In HD160691, however, they allow the system to approach a resonance for
$\sin i \approx 0.5$. }  
\label{fig:genrel} 
\end{figure}

\end{document}